\documentclass[preprint2]{aastex}

\shorttitle{Solar magnetoseismology of jet}
\shortauthors{Morton et. al}

\usepackage{natbib}

\newcommand{\kms}{km\,s$^{-1}$}

\begin{document}
\title{Determination of sub-resolution structure of jet by solar magnetoseismology}

\author{R. J. Morton$^1$, G. Verth$^2$, J. A. McLaughlin$^2$ \& R. Erd\'{e}lyi$^1$}

\affil{$^1$Solar Physics and Space Plasma Research Centre
(SP$^2$RC), University of Sheffield, Hicks Building, Hounsfield
Road, Sheffield S3 7RH, UK\\
$^2$ School of Computing, Engineering \& Information Sciences, Northumbria University, Newcastle Upon Tyne,
NE1 8ST, UK\\ email:r.j.morton@sheffield.ac.uk,
gary.verth@northumbria.ac.uk}

\begin{abstract}
A thin dark thread is observed in a UV/EUV solar jet
in the $171$~{\AA}, $193$~{\AA} and $211$~{\AA} and partially in
$304$~{\AA}. The dark thread appears to originate in the
chromosphere but its temperature does not appear to lie within the passbands of the
Atmospheric Imaging Assembly (AIA) onboard the Solar Dynamics Observatory (SDO). We therefore implement solar magnetoseismology to
estimate the plasma parameters of the dark thread. A propagating
fast kink (transverse) wave is observed to travel along the dark
thread. The wave is tracked over a range of $\sim7000$~km by placing
multiple slits along the axis of the dark thread. The phase speed
and amplitude of the wave are estimated and magnetoseismological
theory is employed to determine the plasma parameters. We are able
to estimate the plasma temperature, density gradient, magnetic field
gradient and sub-resolution expansion of the dark thread. The dark
thread is found to be cool, $T\lesssim3\times10^4$, with both strong
density and magnetic field gradients. The expansion of the flux tube
along its length is $\sim300-400$~km.
\end{abstract}

\maketitle

\section{Introduction}
Short-lived, transient phenomena are commonplace in the solar
atmosphere. In particular, large jet like features are commonly seen
to erupt from the lower to mid solar atmosphere, ejecting material
into the corona. These phenomena are grouped under the term solar
jets, which covers a wide range of explosive events that have been
observed over the last century, including H$\alpha$ surges
\citep{NEW1934}, macro-spicules \citep{BOHetal1975}, UV/EUV
\citep{BRUBAR1983}, X-ray jets \citep{SHIetal1992} and solar
tornadoes \citep{PIKMAS1998}.

It is believed that these events are driven by relatively
localized reconnection events
\citep[e.g.][]{YOKSHI1995,YOKSHI1996, HARR1999}, where emerging magnetic flux reconnects with
existing fields. One feature that has been reported
a number of times among the jets seen in UV/EUV is a large-scale torsional/helical motion
\citep[e.g.][]{SHIMetal1996,PIKMAS1998,PATetal2008,LIUetal2009, KAMetal2010}.
Proposed reconnection models suggest that closed twisted field lines reconnect
with surrounding open field lines \citep[e.g.][]{SHIUCH1985,PARetal2009}.

Whatever the exact excitation mechanism of these jets is, it appears
to coincide with the excitation of a number of different types of
wave phenomena. {Firstly, there is the large-scale torsional/helical
motion reported in both observations \citep[e.g.][]{KAMetal2010} and
simulations \citep[e.g.][]{PARetal2009}, whose interpretation is
still in debate (the candidates being torsional Alfv\'{e}n waves
\citep[e.g.][]{PARetal2009}, or helical kink waves, M.S. Ruderman -
private communication). The second type of reported wave phenomena
is the transverse (fast kink) wave \citep{CIRetal2007,LIUetal2009}.
However, in the majority of observations of transverse (fast kink)
waves in jets, the wave phenomena, in our view, was wrongly
interpreted as an Alfv\'{e}n wave, whose features are quite distinct
from the transverse kink wave \citep{ERDFED2007, VANetal2008b}. The
transverse kink waves are normally reported in a bright, fast moving
part of the jet which is
common to the X-ray jets.} 

The study of wave phenomena in the solar atmosphere proves useful as
demonstrated by the new and expanding field of solar
magnetoseismology, which exploits observed wave phenomena to
determine hard to measure or unmeasurable plasma parameters
\citep{UCHIDA1970, ROBetal1984, ERD2006b,ERD2006}. Recent advances in observing technology
has demonstrated the ubiquitous nature of waves in the solar
atmosphere \citep[see e.g.][for
reviews of waves in the solar atmosphere]{BANETAL2007, ANDetal2009, DEM2009, RUDERD2009, TARERD2009}. However,
magnetoseismology has largely been confined to observations of
structures with relatively long life times, e.g. coronal loops,
\citep[e.g.][]{TARBRA2008,VERERDJES2008,TARERD2009,MORERD2009,MORERD2010,VERTHetal2010,TERetal2011},
prominences \citep[e.g.][]{PINetal2008,SOLetal2010},
although recent papers have extended solar magnetoseismology to transient
events such as X-ray jets \citep{VAGetal2009}, coronal waves
\citep{WESetal2011} and spicules \citep{ZAQ2009,VERetal2011}.

Why should magnetoseismology be used to study solar jets? One reason
is that jets are short lived events with even shorter lived
components, e.g. associated spicule-like features at the footpoints
\citep{STEetal2010}. Regular methods of determining plasma
properties, e.g., magnetic field gradients by spectropolarimetry,
can require fairly long integration times. For example,
\citet{CENetal2010} attempted to measure spicule magnetic field
strengths. However, large signal to noise ratios in off-limb
measurements meant integration times of 45 minutes. Spicules exist
only for 5-15 minutes \citep{ZAQ2009} so such measurements are much
longer than the lifetimes of the objects they are trying to study.
Another reason is that properties of dark features, i.e., little or
no intensity at certain wavelengths, cannot be measured from the
spectral information at those wavelengths. Solar magnetoseismology
negates these difficulties as the changing properties of the wave,
i.e. period, phase speed, amplitude can reveal the required
information on  plasma stratification properties, i.e. magnetic
field gradient, density scale heights, etc. This approach has been
demonstrated successfully by \citet{VERetal2011} in the study of
spicules. Further to this, solar jets are multi-thermal
\citep{KAMetal2010}. X-ray jets in particular have a rapid, hot
ejection of material seen in X-rays and a much cooler ejection of
material that makes up the jet observed in EUV lines \citep[see,
e.g.,][]{KAMetal2010}. Obtaining values of plasma parameters, e.g.,
temperature, and variation of plasma parameters with height, e.g.,
density, magnetic field, of multi-temperature features from the
spectroscopic data available in the narrowband filters, such as
those onboard the Solar Dynamics Observatory (SDO), proves difficult
if not impossible. Solar magnetoseismology, again, negates this
problem. For example, magnetoseismology can be used to determine the
density and magnetic field gradients, and temperature, in theory, of
a feature that appears dark in an SDO filter. Little or no
information could be obtained about the plasma in the object of
interest from the filter response itself. Finally, magnetoseismology
allows the determination of fine structure at sub-resolution scales.
It is at these small scales that wave heating may occur, so the
study of such length scales is vitally important.

In the following we present the observations of a propagating kink
wave observed in a structure associated with a solar (UV/EUV) jet.
The structure is a dark inclusion, in a relatively hot jet, whose
presence appears not to have been seen before (at least in EUV
lines). We then use magnetoseismology to estimate the temperature,
the density stratification and expansion of the observed magnetic structure.
We compare the obtained results with the measured features of the
larger jet structure, magnetic field extrapolations for the quite
Sun magnetic field and chromospheric estimates. 

\begin{figure*}
\centering
\includegraphics[scale=0.85]{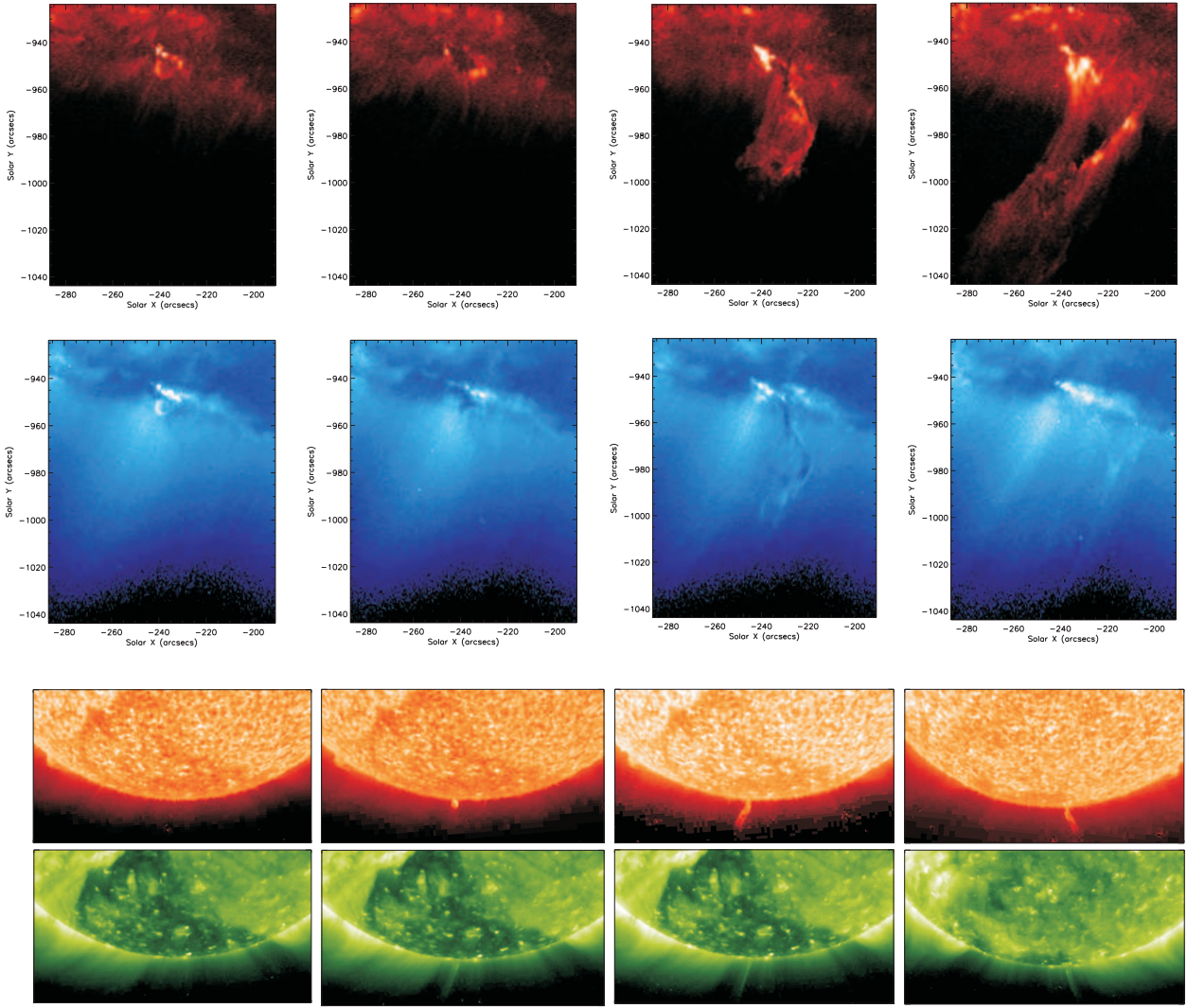}%
\caption{\textit{First and second rows}: Images showing the
morphology of a large jet (observed on $20$ January $2011$) over its
life time taken by SDO/AIA. The panels in the top row are from the
$304$~{\AA} channel taken at 09:16, 09:18, 09:25 and 09:33. The
images on the bottom row are from the $171$~{\AA} channel taken at
the same times. The bright loop is seen in both channels at 09:16,
suggestive of reconnection. A dark feature is seen at 09:18 at the
same position as the bright loop appeared. This dark feature grows
into the dark thread seen in $171$~{\AA} at 09:25. \textit{Third and
fourth rows:} Context images of the jet from STEREO. The first three
panels in each row are from STEREO A and the last is from STEREO B.
The top row are images from the EUVI $304$~{\AA} channel and are
taken at 09:06, 09:26, 09:36 and 09:36. The bottom row are images
from the EUVI $171$~{\AA} channel and are taken at 09:03, 09:28,
09:33 and 09:33. }\label{fig:morph}
\end{figure*}

\section{Observations}
\subsection{Solar jet}
The observations began at {09:00~UT} on the $20$ January $2011$ {and
last an hour and a quarter till {10:15~UT}}, on the south west limb
using the Atmospheric Imaging Assembly (AIA) \citep{LEMetal2011}
onboard SDO which has a spatial resolution of $\sim0.6$~arcsec per
pixel or two-pixel spatial resolution of $\sim870$~km and a cadence
of $\sim12$~s. The solar jet was observed mainly in the $304$~{\AA}
spectral line with faint emission signatures in $171$ {\AA}, $193$
{\AA} and $211$ {\AA}, close to the southern polar coronal hole. The
time series were obtained from the SSW cutout service which provides
level 1.5 data that has already been corrected for flat-field,
despiked and the images co-aligned. The line in which the jet
appears brightest is the $304$ {\AA} which has a peak temperature of
$\sim10^{4.7}$~K (see first row in Figure~\ref{fig:morph}). Our main
interest lies in a dark thread in the $171$ {\AA} line ({second row
in Figure~\ref{fig:morph}}); however, we will also describe the part
of the jet seen in the $304$ {\AA} channel as it will be important
later.

In Figure~\ref{fig:morph} we also provide images from the Sun Earth
Connection Coronal and Heliospheric Investigation (SECCHI)
telescopes onboard the Solar Terrestrial Relations Observatory
(STEREO) \citep{HOWetal2008} to show the orientation of the jet in
the line of sight with respect to SDO (third and fourth rows). The
bright UV/EUV portion of the jet has a $\sim18$ degree angle with
the normal to the Sun's surface; however, it is not possible to
determine the angle between the dark thread and this normal. If it
is similar to the UV/EUV jet then any projection effects should may
be relatively small but should be kept in mind. There is also a
clear emission in the $171$~{\AA} filter from the jet, which appears
as a relatively distinct spike in the quiet corona. The EUV emission
is relatively dim suggesting the bulk of the material in the jet is
less than $1$~MK, although due to blending of the $171$~{\AA} with
cooler lines the actual value could be significantly less than this
\citep[see, e.g.][]{DELMAS2003}.

The jet occurs close to the limb at the southern pole of the Sun and
is situated just outside a coronal hole in a relatively quiet
region. {We do not go into a detailed discussion here of how the jet
may be driven; however, one possible scenario is the jet could be
generated by a reconnection event in the lower solar atmosphere.
This will be the subject of a further investigation.} In
Figure~\ref{fig:morph} we show the morphology of the jet over its
lifetime at various times.

\begin{figure*}[htp]
\centering
\includegraphics[scale=0.8]%
{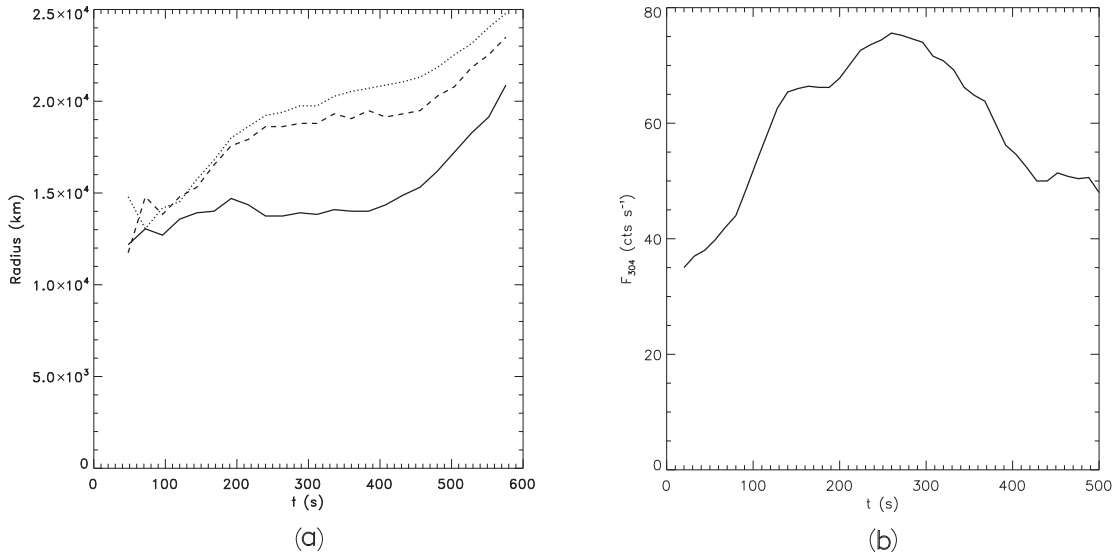} \caption{(a) Expansion of the jet width in $304$~{\AA}
as a function of time plotted for different heights along the jet.
The solid line is taken at $11$~Mm from the base, the dashed line
$14$~Mm and the dotted line $15$~Mm. The time is given in seconds
from $09:22$~UT. (b) Average {intensity of the jet in a cross-cut
taken at a height of 10~Mm above the solar surface} as a function of
time in $304$~{\AA}. The flux is measured in terms of the number of
photon counts per second.}\label{fig:expan}
\end{figure*}

In the $304$~{\AA} line we see a small bright collimated jet rise
from the surface. The jet expands with height and time and during
its main phase, highlights a funnel shape. We place {cross-cuts of
70 pixels in length and one pixel in width} perpendicular to the
central axis of the jet at heights of $\sim10$~Mm, $\sim14.3$~Mm and
$\sim15.7$~Mm above the limb and {use linear interpolation to obtain
an $x-t$ data set}. We then measure the width expansion of the jet
over time at various heights as a function of time,
Figure~\ref{fig:expan}a. (N.B. in all figures from now on the times
are given in seconds from 9:22~UT, which is when the kink wave is
first observed. The benefit of this is clear in Section 4 on
magnetoseismology.) The jet is clearly defined due to the jet
material having a much greater intensity than the surrounding
chromospheric material and the jet edges are determined by eye. {We
suggest there could be an error of $\leq10\%$ on these
measurements}. Figure~\ref{fig:expan}a shows that there is a similar
trend for the expansion of the jet over time at all heights. As
{certain reconnection scenarios involving reconnection between
emerging and existing field describe an untwisting magnetic field},
it is possible that the expansion of the jet cross-section in time
is partly due to to the untwisting of the field. The expansion of
the flux tube in Figure~\ref{fig:expan}a is not constant but goes
through a number of phases.

At around $16$~Mm above the surface the jet experiences a whip-like
effect (see Figure~\ref{fig:morph}) similar to that reported in
\citet{LIUetal2009}. The plasma above the whip appears collimated.
Once the plasma has finished being ejected into the atmosphere, {a
portion of it} returns to the surface. This falling material
highlights the magnetic field and it follows a collimated path back
to the chromosphere suggesting the complete untwisting of the
magnetic field has occurred. The funnel shape of the magnetic field
has now evolved into a column.

Further to this, {we calculate the median value of intensity flux,
$F$, in the $304$~{\AA} channel of the pixels in the cross-cut
placed at $10$~Mm, which were determined to lie within the jet
boundaries. This is plotted as a function of time in
Figure~\ref{fig:expan}b.} We do this because the intensity is
dependent upon thermodynamic quantities, e.g., temperature and
density, so provides an estimate of how the density/temperature in
the jet evolves as a function of time.

In the $171$~{\AA} channel the jet appears completely different to
that seen in $304$~{\AA}. At 09:16 a bright loop is clearly seen
with the eruption starting two minutes after. A thin, dark thread,
with a width at resolution scales, is then seen to be ejected into
the atmosphere along with a very small amount of bright material.
The dark material appears to come from the chromosphere, appearing
(apparently) from behind the bright loop (09:18)
(Figure~\ref{fig:morph}). This is the first appearance of the dark
thread that is shown in the image at 09:25. The dark thread is
co-spatial and co-temporal with the UV/EUV component of the jet in
the $304$~{\AA} channel but is barely visible amongst the hotter
plasma. In the middle right panel of the $304$~{\AA} images a
bright, thin thread of plasma is seen to highlight the right hand
jet edge. This can just be made out in the corresponding $171$~{\AA}
image running along the righthand side of the dark thread. Towards
the base of the bright thread in the $304$~{\AA}, a small dark
inclusion can be seen which is the dark thread (e.g. third panel,
top row, Figure~\ref{fig:morph}). This suggests that the thin, dark
thread is a flux tube (or flux tubes) contained within a large
collection of flux tubes, i.e. the jet seen in the $304$~{\AA}
channel.  The dark thread survives for around 10 minutes before the
material either traveling higher into the atmosphere or returning to
the surface. Further, the dark thread jet has an angle of $\sim35$
degrees to the normal from the solar surface. {The plane in which
\emph{this} angle is measured, i.e., the plane of view of SDO/AIA,
is almost perpendicular to the plane of view of STEREO. Hence, the
angle measured for the bright part of the jet in STEREO is measured
in a different plane to this angle.}

Aside from the dark thread, only a small amount of bright plasma is
seen to be ejected into the atmosphere in $171$~{\AA}. This suggests
that the density and/or temperature of the dark thread plasma must
be different from the material seen in the $304$~{\AA} channel. Due
to its appearance as a dark inclusion we suggest the material is
much cooler and/or denser (hence, optically thicker) than the other
jet material. The peak sensitivity of the $304$~{\AA} channel is
$T\sim10^{4.7}$~K, so the material in the dark thread must be at
least $T<5\times10^4$~K. This, along with the appearance of the
bright loop at the jet footpoint, supports the popular conjecture
that the reconnection event responsible for the jet occurred at
chromospheric heights. The dark thread could be similar to the cool
chromospheric jet observed by \citet{LIUetal2009}. One difference
between the jet observed here and that in \citet{LIUetal2009} is
that the ejected material in \citet{LIUetal2009} returns to the
surface along a similar path to that of its ejection. Here, some of
the material is ejected higher into the atmosphere while the rest
returns to the surface along a different path at least a few Mm from
the ejection site.

The dark thread has a width close to the resolution limit so the
expansion that is measured for the UV/EUV jet structure is not seen
in the dark thread. The outflow of the dark thread material is also
a lot slower than outflow of bright material seen in the
$171$~{\AA}. In $171$~{\AA} we can measure the speed of the outflow
for the dark thread by using two methods. For the initial rise of
the dark feature we use a running difference movie to track the dark
thread front and obtain average rise speeds of $63$~\kms. After the
initial rise we employ a method of contouring features of different
intensity and tracking the intensity features through the images
\citep[see, e.g.][for details]{WINetal2001}. {The flow in the dark
thread is given as an average speed along the thread of the
different intensity features identified and tracked through the
thread. Here, we term the initial rise or possible \lq{extension}\rq
of the dark thread as outflow. In the following section we need to
subtract any apparent motion, both flow and extension, from measured
phase speeds. We use the term outflow to describe the observed
upwards motion of plasma, making no definite distinction between
possible extension or flow.}

\begin{figure}
\centering
\includegraphics[scale=0.4]%
{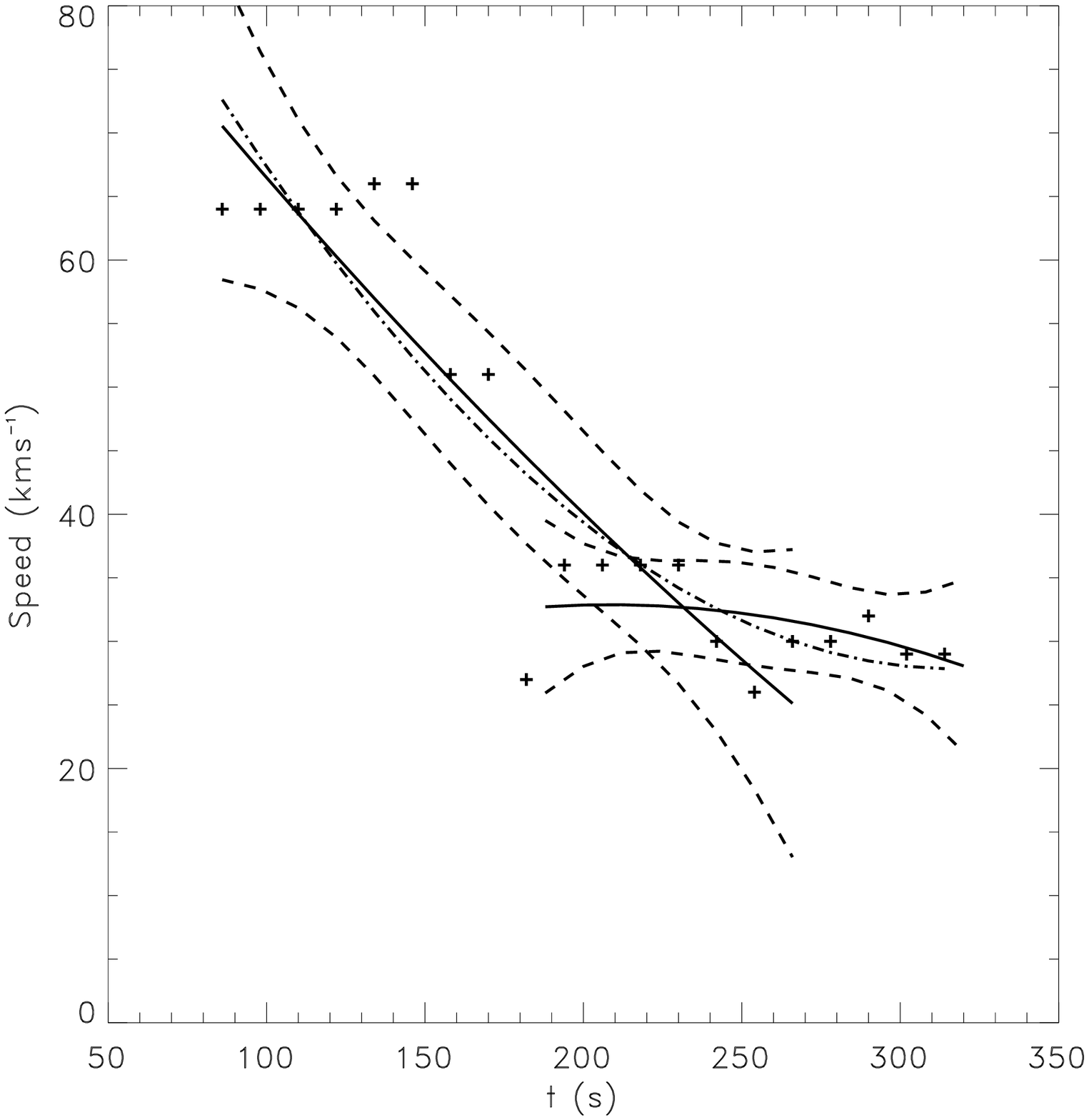} \caption{Measured values of flow as a function of
time along the dark thread in the $171$~{\AA}. The crosses are the
measured values while the solid lines correspond to the quadratic
fits to {the measured values smoothed with 5 point boxcar average}.
The dashed lines are the $95$~\% confidence levels from the fits.
The dashed-dot line is a quadratic fit to all the data points
}\label{fig:flow}
\end{figure}

In Figure~\ref{fig:flow} we show the outflow speeds along the dark
thread as a function of time. The speed of the outflow lies within
the range of outflow speeds reported in \citet{LIUetal2009}. We then
smooth the measured values of flow {with a 5-point boxcar average}
and fit a quadratic profile to the smoothed points which is the dash
dot line in Figure~\ref{fig:flow}. It would appear that there are
two distinct phases to the flow; a phase with decreasing flow speed
and a phase of almost constant flow speed. Separating these into two
different phases we construct quadratic fits to the two phases
(solid lines). The second phase can be seen to have an almost
constant flow speed. It is unknown why the outflow rate appears to
have these two distinct phases, but is probably connected to the
nature of the explosive (reconnection) event.

\subsection{Transverse wave}
We can determine very little about the dark thread from the SDO
filters due to its low intensity in all SDO filters. However, we
observe a propagating transverse wave traveling along the thread.
The natural interpretation for the wave is a propagating transverse
wave (propagating kink wave). A number of previous authors have
reported observed displacements of the jet axis as the Alfv\'{e}n
wave. However, Alfv\'{e}n waves are oscillations of constant
magnetic surfaces and do not displace the flux tube axis, the only
mode that causes a visible displacement of the flux tube axis is the
fast kink mode \citep[see, e.g.][for discussion]{ERDFED2007,VANetal2008b}. The
transverse wave should allow us to estimate certain plasma
parameters by means of solar magnetoseismology. First, in this subsection we
describe how we obtain the information about the wave and in the
following sections we discuss the magnetoseismology.

\begin{figure}
\centering
\includegraphics[scale=0.5]%
{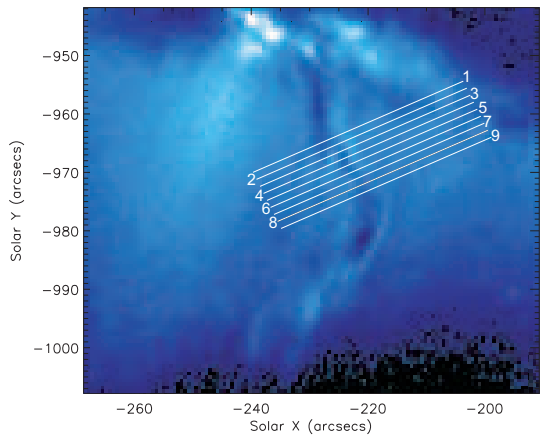} \caption{Image of the dark thread showing the
cross-cut positions. The observed transverse motion propagates from
cross-cut 1 to 9.}\label{fig:slits}
\end{figure}

\begin{figure*}
\centering
\includegraphics[scale=0.8, clip=true, viewport=0.0cm 0.0cm 25.0cm 17.5cm]%
{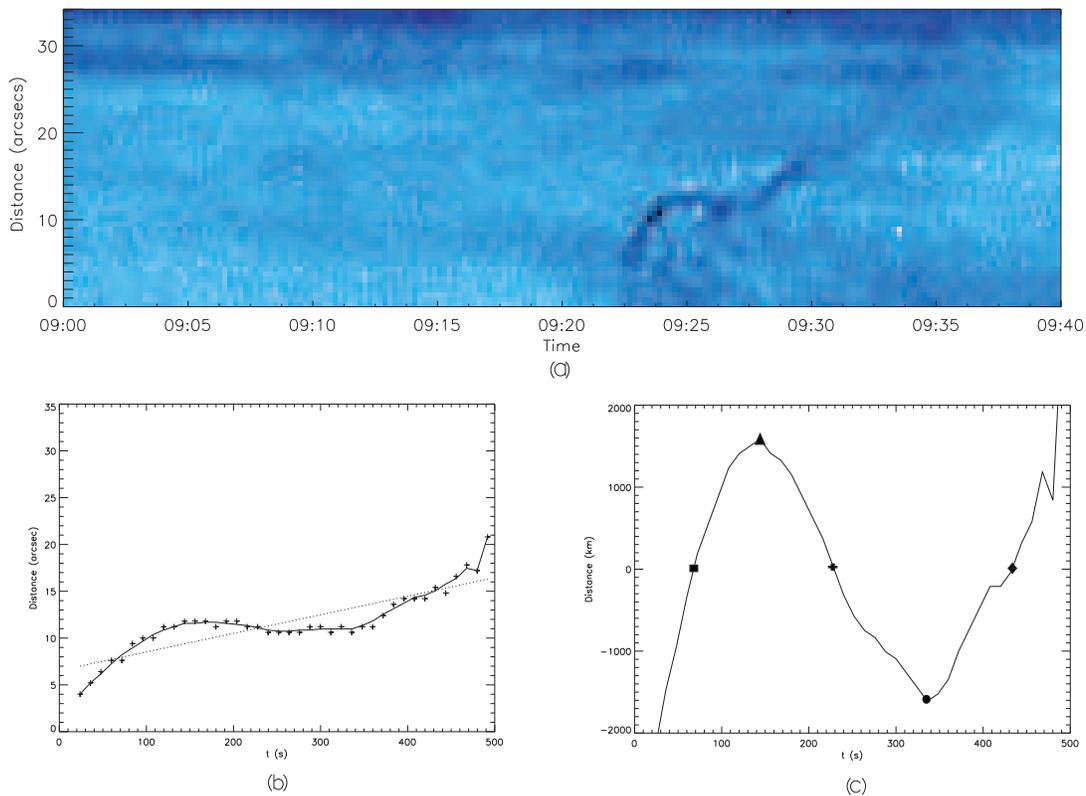} \caption{(a) A sample cross-cut $x$-$t$ plot
(cross-cut No.3). The transversal kink motion of the dark thread is
clearly seen. (b) and (c) An example of the fitting procedure for
the third cross-cut. (b) The middle panel shows the data points
(crosses) over plotted with a smoothed profile (solid line). The
dotted line shows the linear fit to the wave. (c) The right hand
panel shows the wave with the linear fit subtracted. The symbols
indicate the first zero-crossing (\textit{square}), the maximum
(\textit{triangle}), the second zero crossing (\textit{cross}), the
minimum (\textit{circle}) and the third zero crossing
(\textit{diamond}).}\label{fig:fit_smooth}
\end{figure*}

We determine the line parallel to the final axis of the dark thread
and then place $9$ cross-cuts ({linearly interpolated})
perpendicular to the line, with the cross-cuts separated by
$0.97$~Mm (see, Figure~\ref{fig:slits}). The wave appears at
approximately half way along the thread, which is approximately
$\sim7$~Mm above the limb in $171$~{\AA}. We do not consider
projection effects along the line of sight for estimates on
distances and heights. The angle between the normal to the solar
surface and the UV/EUV jet is approximately $\sim18$ degrees (see,
Figure~\ref{fig:morph}), so the correction will only be relatively
small, $\sim5$~\% on measured distances. The propagating kink wave
can be seen up to a height of $\sim15$~Mm in the atmosphere. This
corresponds to the wave traveling a distance of $\sim11.6$~Mm along
the dark thread. An example of the oscillation observed in one of
the cross-cuts is shown in an $x$-$t$ plot in
Figure~\ref{fig:fit_smooth}a.

\begin{figure*}
\centering
\includegraphics[scale=0.8]%
{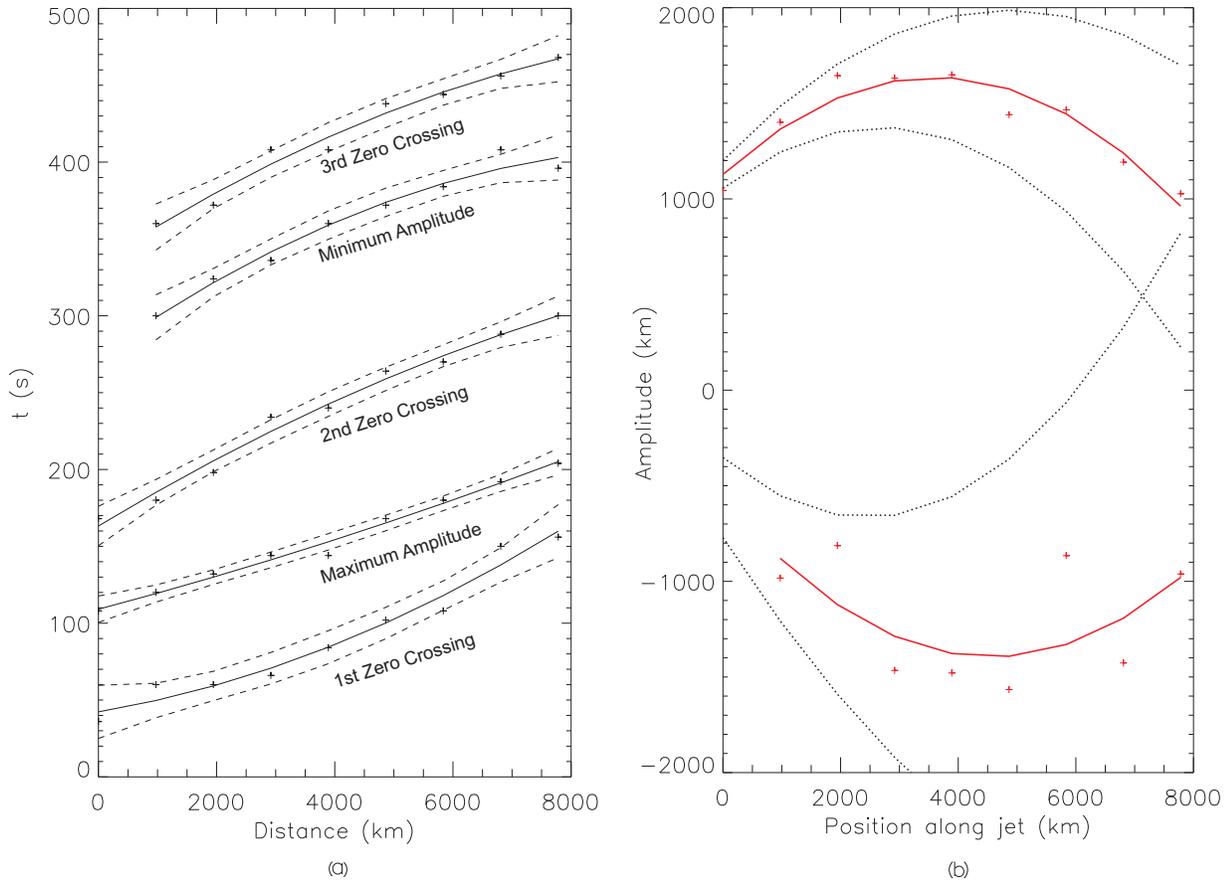} \caption{(a) Shown are the times at which the
zero-crossings and maximum and minimum amplitudes appear at the
different heights (cross-cuts). The data points (crosses) have been
fitted with a quadratic profile (solid line) and the $95\%$
confidence interval is shown for the fit. (b) Maximum and minimum
values of the amplitude envelope plotted as a function of height.
The crosses correspond to the measured values, the solid line is a
quadratic fit to the data points and the dashed lines correspond to
a sigma confidence interval for fit to the maximum amplitude
envelope and 0.5 sigma confidence interval for fit to the minimum
amplitude envelope. }\label{fig:data_points}
\end{figure*}

\begin{figure*}
\centering
\includegraphics[scale=0.8]%
{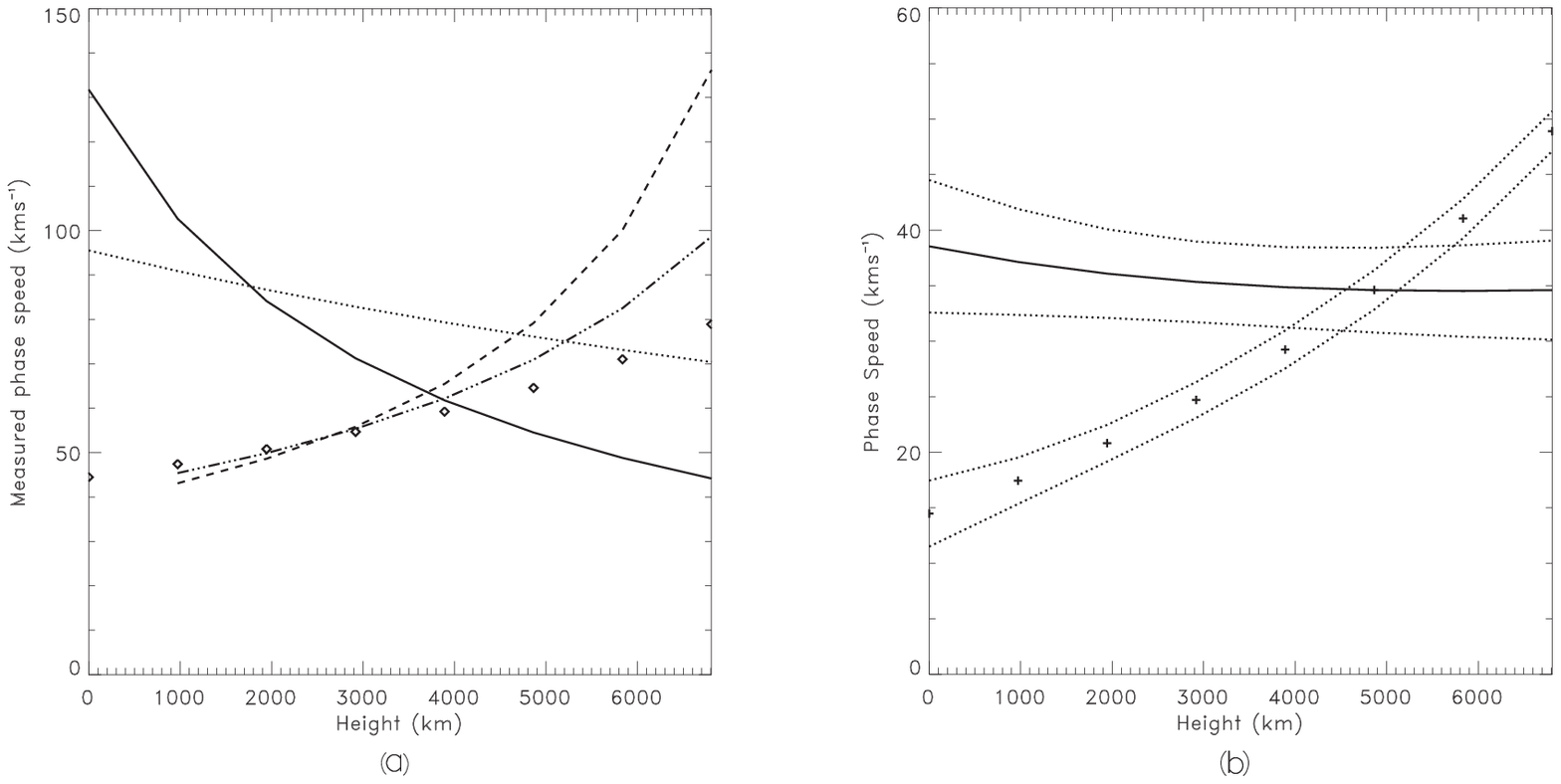}
\caption{(a) Phase speeds for first zero (solid),
maximum amplitude (dotted), second zero (diamonds), minimum
amplitude (dashed), third zero (dashed dot). (b) Measured phase
speed minus the measured flow for the maximum amplitude (solid line)
and second zero crossing (crosses). The dashed lines correspond to
the sigma value error on the phase speeds. }\label{fig:phase_speed}
\end{figure*}

To obtain information about the wave we determine the pixel in each
time frame of the $x$-$t$ plot that has the minimum intensity. The
dark thread has an almost inverse gaussian intensity profile across
its width and the minimum intensity should correspond closely to the
minimum of this inverse gaussian. We then smooth the data points of
the oscillation using a 5 point box-car function before applying a
linear fit to the smoothed data. The linear fit is subtracted from
the smoothed data to remove the trend. An example of these steps is
demonstrated in Figure~\ref{fig:fit_smooth}b and c. We then
determine the times at which the zero crossings and the maximum and
minimum values of amplitudes occur in the different cross-cuts. The
magnitude of the maximum and minimum values of the amplitudes are
also determined. Above the $9$th cross-cut the strength of the wave
signal is weak making it difficult to obtain fits to the observed
profile.
We begin by plotting the travel times at which the zero crossings
and the maximum and minimum values of amplitudes occur at the
different heights in Figure~\ref{fig:data_points}a. {The cross-cut
labelled No.~$1$ in Figure~\ref{fig:slits} is the first cross-cut in
which we see the oscillation; hence, in
Figure~\ref{fig:data_points}a we measure distance from this
cross-cut.} The crosses correspond to the values taken from the
observations and the lines correspond to a quadratic fit to the data
points of each feature. The gradient of the quadratic fits gives the
speeds at which the zero crossings and the maximum and minimum
values of amplitudes travel along the dark thread and is shown in
Figure~\ref{fig:phase_speed}a.

One might expect that the speeds of each point should have similar
behavior as they propagate to the different heights. However, as we
have illustrated in Figures~\ref{fig:expan} and \ref{fig:flow}, the
plasma properties, i.e. flow, magnetic field and density, in the
larger collection of flux tubes ($304$~{\AA} jet) and the dark
thread are subject to large changes over the lifetime of the
jet/thread. If we assume a linear fit so that the phase speeds are
constant, we see the trend is a decrease in the phase speed over
time. This appears natural as the phase speed of the transverse wave
is proportional to the magnetic field, so an untwisting field would
decrease the field strength, hence, the value of phase speed will
also decrease. N.B. Neglecting inclination of the flux tube with
respect to the normal to the solar surface gives errors of
$\pm3$~kms$^{-1}$ which lie well within the errors/confidence levels
given by the fitting procedure.

Finally, we plot the size of the maximum and minimum amplitude
envelopes in Figure~\ref{fig:data_points}b. It is seen from the
amplitude envelope plot initially there is an increase in amplitude
with height followed by a decrease. The confidence intervals on the
fits demonstrate that there is a possibility that the amplitude does
not necessarily have to be decreasing after $4000$~km. It is clear
from the confidence intervals in Figure~\ref{fig:data_points}b that
the minimum amplitude appears to have large errors associated with
it. In the following sections we will use only the maximum amplitude
envelope to increase confidence in our results.

\section{Theory}
Before we analyze the wave motion any further we need to determine
which, if any, theory already developed is relevant to these
observations. A complete theory that would describe kink motion in
the jet scenario has not yet been fully developed due to the immense
complication associated with including all the topological and
dynamic features that the reported jet possess. However, we will
assume static behavior and adapt the theory for a magnetic flux tube
with both magnetic and density stratification
\citep{VERERD2008,RUDetal2008}. If the plasma is indeed cool as we
suspect, then it may be better to use models including a partially
ionized plasma, e.g., \citet{SOLetal2009,SOLetal2009b} who model
prominence plasmas. {However, the results from these models indicate
partial ionization, at least in single fluid theory, has a small
effect on the propagation and damping of fast kink oscillations (for
typical observed wavenumbers).}

First, we need to make sure any assumptions used in the derivations
of the previous theories are also adequate for this dark thread: (i)
The thin tube approximation applies. For this we require the
wavelength of the oscillation to be much greater than the radius.
The width of the dark thread is measured to be $<500$~km.  We can
estimate the wavelength/wavenumber of the kink wave. The typical
period of the oscillations (e.g. see oscillation in
Figure~\ref{fig:fit_smooth}) is $360$~s. Using
$P=2\pi/\omega=\lambda/c_k$ and values of phase speed from
Figure~\ref{fig:phase_speed}b then, $5400<\lambda<18000$~km and
hence $R/\lambda\lesssim0.1$. (ii) The plasma-$\beta$ is small.
Estimates for the sound speed in the chromosphere are in the range
$5-12$~kms$^{-1}$ ($T=5\times10^4-1\times10^5$~K) while the
Alfv\'{e}n speed is between $50-200$~~kms$^{-1}$. The plasma-$\beta$
is given by $\beta=2c_s^2/(\gamma v_A^2)$ so estimates for the jet
give $\beta<0.25$. From these estimates we can consider the
plasma-$\beta$ in the dark thread to be small. (iii) The plasma
inside the dark thread is much denser than the surrounding material,
i.e. $\rho_i>>\rho_e$. Although this is not a necessary condition,
it allows simplification of the governing equations and suggests
that the properties of the internal plasma dominate the wave
behavior. The dark thread appears to originate from the
chromosphere/transition region and ejects the material from these
regions high into the corona. It is generally well accepted that
corona is much less dense than the lower layers of the solar
atmosphere; hence, $\rho_i$ is at chromospheric/transition region
densities and $\rho_e$ is at coronal densities. However, the plasma
at UV/EUV temperatures in $304$~{\AA} appears to surround the dark
thread, so the external plasma could be considered to be at upper
chromosphere/transition region temperatures and densities. It is
then much less clear whether $\rho_e$ is less than $\rho_i$.

First, the phase speed for a fast kink (or transverse) mode in a
thin tube is given by
\begin{equation}\label{eq:kink}
c_k(z,t)=\left(\frac{{B_i(z,t)^2+B_e(z,t)^2}}{{\mu_0(\rho_i(z,t)+\rho_e(z,t))}}\right)^{1/2},
\end{equation}
where the subscripts $i$ and $e$ refer to the internal and external
values of the plasma quantities. Instead of making any assumptions
about the relative magnitudes of the densities and magnetic field we
will rewrite this in terms of the average density and magnetic field
\begin{equation}
c_k=\frac{B}{\sqrt{\mu_0\rho}},
\end{equation}
where $\rho=(\rho_i(z,t)+\rho_e(z,t))/2$ and
$B^2=(B_i(z,t)^2+B_e(z,t)^2)/2$.

Now, in \citet{RUDetal2008} the kink wave equation for a
magnetostatic flux tube with both magnetic and plasma density
stratification in the longitudinal direction is given by
\begin{equation}\label{eq:expand}
\frac{d^2\eta}{dz^2}+\frac{\omega^2}{c_k^2(z)}\eta=0,
\end{equation}
where $\eta=\xi_\bot/R(z)$, $\xi_\bot$ is the displacement
perpendicular to the magnetic field, $R(z)$ is the loop radius and
$\omega$ is the frequency of the transverse oscillation.



First, we introduced a scaled length, $\zeta=\epsilon z$ under the
assumption the length scale of the changes in the background plasma
parameters, i.e. density, magnetic field, are small compared to the
wavelength, hence, the kink speed changes slowly along a stratified
and expanding waveguide. We now apply the WKB approximation
\citep[see e.g.,][]{BENORS1978} in the $z$ direction, assuming that
\begin{equation}\label{eq:wkb}
\eta=A(\zeta)\exp(-i\epsilon^{-1}\Theta(\zeta)).
\end{equation}
Substituting Equation~(\ref{eq:wkb}) into (\ref{eq:expand}) and
taking the {zeroth} order terms, with respect to $\epsilon$, gives
the frequency
\begin{equation}
\omega=c_k(z)k,
\end{equation}
where $k=d\Theta/d\zeta$. The {first} order terms provides an
equation for the amplitude of the wave, namely
\begin{equation}\label{eq:amp}
\frac{A'}{A}=-\frac{k'}{2k},
\end{equation}
where the dash corresponds to derivative with respect to $\zeta$.
Solving this and using our definition of $\eta$ we obtain
\begin{equation}\label{eq:main}
\xi_\bot=\sqrt{\frac{c_k(z)}{\omega}}R(z).
\end{equation}
The frequency is independent of the changes in the spatial
direction, hence, if we can determine the change in phase speed and
spatial displacement of the thread (amplitude) with height then the
expansion along the flux tube can be calculated.


\section{Solar magnetoseismology}
Now, the data we have determined from the observations contains a
large amount of information. To demonstrate the magnetoseismology
here we shall only use two of the obtained phase travel times, for
which the reasons will become clear.

First, we return to the plot of the phase speeds shown in
Figure~\ref{fig:phase_speed}a. For a propagating wave in a flowing
medium, the measured phase speed will be a sum of the kink speed and
the flow speed (if the wave is propagating in the same direction as
the flow). If we subtract the measured value of flow from the
measured phase speeds then we see that the actual phase speeds cover
a much smaller range of values (e.g. see
Figure~\ref{fig:phase_speed}b). From Equation~(\ref{eq:kink}) we can
see the kink speed is also dependent upon the magnetic field and the
density. The dynamic changes observed in the large jet could also be
occurring in the dark thread and could be sufficient to explain the
remaining differences between the phase speeds.

\subsection{Scale height and temperature estimates using the phase speed}
The UV/EUV jet is clearly highly dynamic, where magnetic field and
density undergo rapid changes. However, if we consider
Figure~\ref{fig:expan}, we can see that there is a period of time
when the time-dependent expansion and change in intensity of the jet
is at a minimum ($t\approx170$~s to $t\approx400$~s). In this period
of time the total expansion of the jet is $\lesssim5\%$ and the
change in intensity is $\lesssim10\%$. For this time period, the
second zero crossing is propagating along the thread. We then make
the assumption that the activity in the dark thread is also at a
minimum at this point. If this is the case, the change we see in
phase speed at the different heights is due only to the structuring
along the thread and not due to dynamic behavior. The second zero
crossing appears first at around 156~s (9:24:30 UT), propagates
along the cross-cuts within $110$~s and experiences an increase in
phase speed as it travels along the thread. Using the estimate of
flow speed along the jet during this period (Figure~\ref{fig:flow})
we subtract flow from the measured phase speed to obtain the kink
speed estimate (Figure~\ref{fig:phase_speed}b).

Now, since we have no information on the expansion of the dark thread with
height we make the initial assumption that the magnetic field is
constant with height. This assumption is only for ease of
calculation and provides an upper bound on the density scale height.
As will be demonstrated in the next section, this assumption is not necessarily valid in such solar waveguides.

First, we assume the density is gravitationally stratified, such
that
\begin{equation}
\rho\propto\exp\left(-\frac{z}{H_{eff}}\right),
\end{equation}
where $ H_{eff}={H}/{\cos\theta} $ is the effective density scale
height due to the inclination of the dark tube from the normal to
the surface \citep[see, e.g.][]{ASC2004}, $H$ is the density scale
height and $\theta$ is the inclination angle. We need the
relationship
\begin{equation}\label{eq:kink_rel}
\left(\frac{c_{k0}}{c_{k}}\right)^2=\frac{\rho}{\rho_0}=\exp\left(-\frac{\triangle
z}{H_{eff}}\right),
\end{equation}
where the subscript $0$ refers to the `first' value of phase speed
and density, $\triangle z$ is the difference in height between where
$c_{k0}$ and $c_{k}$ are measured. {Substituting in the obtained
values of phase speed for the second zero crossing with flow
subtracted (Figure~\ref{fig:phase_speed}b), where $c_{k0}$ is the
phase speed at height $0$~km and $c_k$ is at height $7000$~km,} we
obtain that $H_{eff}=2.4\pm0.5$~Mm corresponding to
$T\sim0.05\pm0.01$~MK, where we have used the relationship
\begin{equation}
H=\frac{2k_B}{\mu m_Hg}{T},
\end{equation}
and $k_B$ is the Boltzmann constant, $\mu$ is the mean molecular
mass, $m_H$ is the mass of hydrogen and $g$ is the gravitational
acceleration ($2k_B/\mu m_Hg\approx47$~Mm/{MK}, see e.g.,
\citealp{ASC2004}).

{We can attempt to take into account expansion with height, which
gives a new relationship between phase speed and scale-height,
namely
\begin{equation}
\left(\frac{c_{k0}}{c_{k}}\right)^2=\frac{B_0^2\rho}{B^2\rho_0}=\frac{R^4}{R_0^4}\exp\left(-\frac{\triangle
z}{H_{eff}}\right).
\end{equation}
Using the values of expansion in the measured range for the larger
UV/EUV jet, i.e. $R/R_0\sim1.35$ (from Figure~\ref{fig:expan}a), and
the same values of phase speed for the second zero crossing}, we
obtain $H_{eff}=1.6\pm0.5$~Mm corresponding to
$T\sim0.03\pm0.01$~MK. Any larger value of expansion for the dark
thread will give even smaller values of $H$ and $T$, hence, assuming
no magnetic field expansion along the thread provides an upper bound
for density scale heights and temperature estimates. The obtained
estimates for the temperature correspond to an intermediate
temperature between chromospheric temperatures ($10^4$~K) and
transition region temperatures ($10^5$~K). {Co-spatial emission with
dark thread is observed in the $1600$~{\AA} channel, which has
contributions from transition region elements (i.e. C IV).
Investigation of the emission reveals it appears to be transition
region material associated with, rather than part of, the ejection
of the dark thread. This suggests the material is at least cooler
than the transition region plasma, i.e. $T<10^5$~K.}

As we see signals of a reconnection event, this could suggest that
the thread is a chromospheric loop that has been heated. The
obtained estimates of temperature also agrees with the hypothesis
given by our earlier arguments on why the filament appears dark in
the $171$~{\AA} channel, furthering credence to the suggestion that
we see a similar event to \citet{LIUetal2009}. Further, the obtained
estimates for the density scale height are greater than the
estimated scale heights derived for spicules from magnetoseismology
in \citet{VERetal2011} because the thread is hotter than Ca II H
spicules.


\begin{figure*}
\centering
\includegraphics[scale=0.9]%
{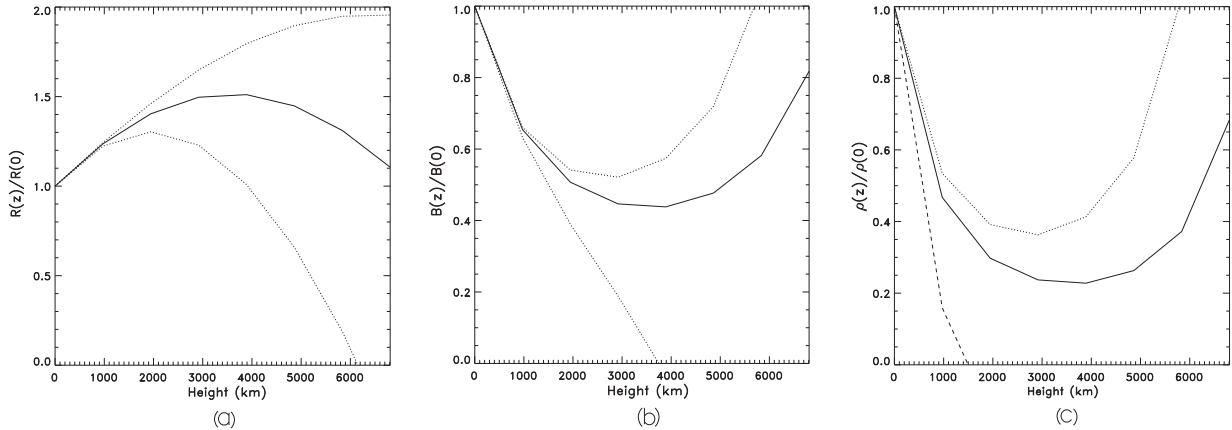} \caption{(a) The estimated expansion of the flux tube
(solid line) along the dark thread derived by means of
magnetoseismology. The dashed lines correspond to the $95\%$
confidence interval. (b) The estimated change of magnetic field
strength (solid line) along the dark thread from magnetoseismology.
The dashed lines correspond to the $95\%$ confidence interval. (c)
The estimated change in density (solid line) along the dark thread
obtained by solar magnetoseismology. The short dashed line
corresponds to the $95\%$ confidence interval, whereas the long
dashed line corresponds to a $70\%$ confidence interval.
}\label{fig:seis_expan}
\end{figure*}

\subsection{Estimation of flux tube expansion}
Now, the rest of the phase travel times occur during times of
significant dynamic behavior in the larger jet (see,
Figure~\ref{fig:expan}). We do not know if the dark thread also
experiences similar dynamic behavior. To make progress, we assume
the thread is at least in a quasi-static state during the time for
which the position of the maximum amplitude of the wave propagates
through the cross-cuts. This means that the following
magnetoseismological results can be interpreted as a time average of
the thread's structure.

The phase travel time for the position of the maximum amplitude has
a very similar fit when either a quadratic or exponential fit is
used. We use an exponential fit since this is particulary convenient
for calculating the uncertainty in the rate of change of phase speed
along the thread. This is necessary for quantifying the uncertainty
in magnetic field strength and plasma density gradients inferred
from the following magnetoseismology.  In
Figure~\ref{fig:phase_speed}a the measured phase speed is shown and
in Figure~\ref{fig:phase_speed}b we plot the measured phase speed
minus the flow, i.e. what should be the actual phase speed of the
wave, for the position of the maximum amplitude.

Using Equation~(\ref{eq:main}) it is obvious that we can write
\begin{equation}
\frac{R}{R_0}=\sqrt{\frac{c_{k0}}{c_k}}\frac{\xi_{\bot}}{\xi_{\bot0}},
\end{equation}
where $\xi_{\bot0}$ and $R_0$ are the values obtained at lowest
height and $\xi_\bot$ and $R$ are the values obtained at a greater
height. Using the measured value of amplitude and the actual phase
speed we can determine the the rate of flux tube expansion.
Figure~\ref{fig:seis_expan}a shows the normalized expansion along
the dark thread.

What likely values of expansion should we expect for the dark
thread? It is possible, as mentioned, that the dark thread will
display the same behavior as the larger UV/EUV jet. The local
expansion with height of the dark thread is initially similar to
that measured in the global jet structure shown in
Figure~\ref{fig:expan}a. The large UV/EUV jet takes on the shape of
a magnetic funnel, suggested as being the standard magnetic
configuration for the quiet Sun \citep[e.g.][]{GAB1976,PET2001}.
Expected expansion factors from extrapolations for quiet Sun
magnetic funnels, e.g. \citet{TANetal2010}, are $4-12$ for heights
less than $20$~Mm and $1-3$ for heights greater than $20$~Mm. The
calculated expansion in Figure~\ref{fig:seis_expan}a also appears to
show the dark thread exhibiting a contraction of the cross-sectional
area, with the profile being almost symmetric about $3500$~km. Is it
possible this is a real contraction? Two likely scenarios suggest
themselves if it is a real contraction: (i) if the thread is a
chromospheric loop and reconnection occurred close to a footpoint of
the loop, then we may expect the dark thread to have a relatively
symmetric profile about the apex; (ii) there are strong magnetic
elements either side of the dark thread forcing the field lines to
converge. However, the results from the previous section suggest the
density is decreasing along the dark thread. It is possible this
contraction has disappeared (due to untwisting of the thread)
between the time the maximum amplitude travels along the thread to
the time the second minimum travels. The decreasing flow speed may
be a sign that the thread has untwisted/opened up significantly
between these two events. However, the minimum amplitude also
appears to show a decrease of amplitude with height.

Another option is that the kink wave experiences damping. Resonant
absorption is known to be an efficient method for the damping of
propagating kink waves \citep[e.g.,][]{TERetal2010c, GOOSSetal2011}
and is an ideal candidate for plasma heating
\citep{ERD1998,TARERD2009}. Also, dynamic behavior has been shown to
attenuate/amplify the amplitudes of MHD waves
\citep{MORERD2009b,MORHOOERD2010, RUD2011, ERDetal2011}. This means
the contraction would be an artifact as the oscillation is damped.
We can estimate the wavelength/wavenumber of the kink wave. The
typical period of the oscillations (e.g. see oscillation in
Figure~\ref{fig:fit_smooth}) is $360$~s. Using $P=2\pi/\omega$ then,
$5400<\lambda<18000$~km and
$5\times10^{-5}<k<25\times10^{-4}$~km$^{-1}$ for
$15<c_k<50$~kms$^{-1}$. This means $0.03<kR<0.15$ for a radius of
$\sim725$~km (the radius is at resolution length scales) and
resonant damping is efficient at these wavelengths.

We can also determine the magnetic and density gradients along the
dark thread. The magnetic field varies as
$$
B(z)\propto\frac{1}{R^2(z)},
$$
and this is shown in Figure~\ref{fig:seis_expan}b. Further, using
the values of the actual phase speed and the relation
$$
\rho\propto\frac{B^2}{c_k^2}
$$
we can determine the density gradient, Figure~\ref{fig:seis_expan}c.
If we fit a decreasing exponential profile to the decreasing section
of the density we obtain a scale height of $H\sim1.2$~Mm and thread
temperature of $T\sim2.6\times10^4$~K. This temperature is less than
the estimates obtained in the previous subsection and adds weight to
the suggestion the thread is chromospheric in origin. Note, the
reason for a $70\%$ confidence interval in
Figure~\ref{fig:seis_expan}c rather than a $95\%$ interval is due
being unable to calculate higher confidence due to large relative
errors on the density.

\begin{figure}
\centering
\includegraphics[scale=0.5]%
{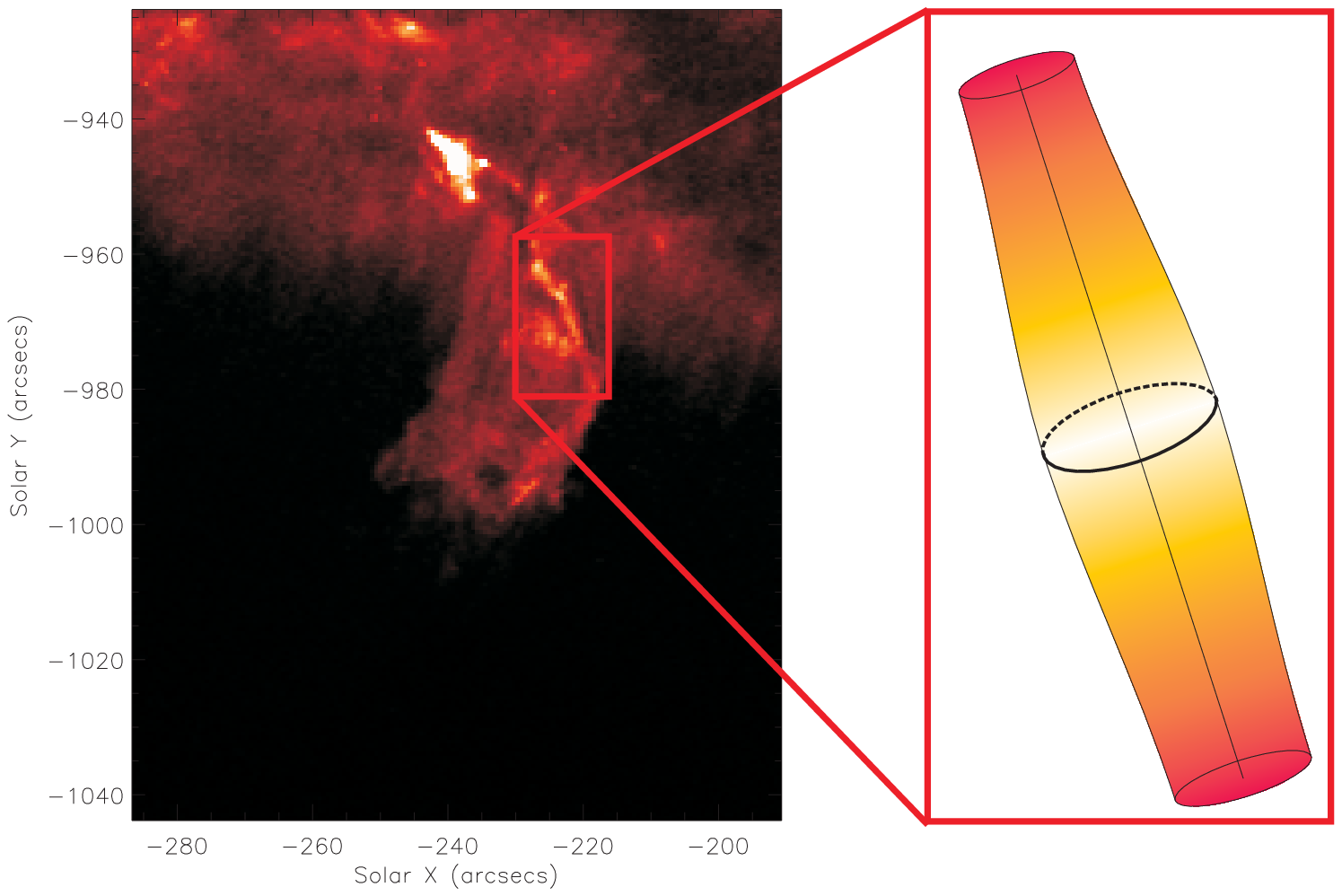} \caption{Schematic representation of the
sub-resolution features of the dark thread determined from
magnetoseismology.}\label{fig:flux_image}
\end{figure}

In Figure~\ref{fig:flux_image} we provide a schematic representation
of the apparent scenario in the jet suggested by the
magnetoseismology. It demonstrates the sub-resolution expansion of
the thread and the density stratification that can be inferred from
such an approach.

\section{Conclusions}
In the present study we demonstrate the successful application
of solar magnetoseismology to a solar jet. In particular we determine
information on various plasma parameters of a dark thread that is
observed to occur with a UV/EUV solar jet. From the
magnetoseismology we are able to estimate the temperature of the
dark thread, the sub-resolution expansion and the magnetic and
density gradients along the thread. It would be difficult to obtain
this information through analysis of the SDO/AIA filter data or
spectral lines profiles, e.g. Stokes profiles, line widths, etc.

The nature of the dark thread is still a mystery as no similar
feature has been reported previously. The temperature estimates
($T\sim3\times10^4$) obtained from the magnetoseismology suggest the
plasma is cooler than the observable temperature ranges of both the
$304$~{\AA} and $171$~{\AA}. This is probably the reason, along with
the plasma being optically thick/dense, that the thread appears as a
dark inclusion in the filters. A dense, $T\sim3\times10^4$ plasma is
likely to have originated in the chromosphere. The thread feature is
possibly part of an emerging flux region that was responsible for
the reconnection that caused the jet. The closest comparison we find
is the chromospheric jet observed by \citet{LIUetal2009,
LIUetal2011}.

With respect to the magnetoseismology, the fact we obtain reasonable
temperature estimates for the plasma gives confidence that the
density and magnetic field gradients also obtained are physical. The
only issue that arises is the apparent contraction with height of
the loop. A reasonable scenario that fits with the entire picture of
the jet and the other results from the seismology is difficult to
come by. It could be due to having neglected any influence of wave
damping (e.g. due to resonant absorption) or because we are missing
information on the dynamic evolution of the dark thread. Further
theoretical studies into the influence of dynamic plasma behavior
on wave propagation are required if a full picture of an objects
dynamic behavior is to be found via magnetoseismology alone.

\begin{acknowledgements}
The authors thank A. Engell, D. Jess, and P. Grigis for a number of
useful discussions. RE acknowledges M. K\'eray for patient
encouragement. The authors are also grateful to NSF, Hungary (OTKA,
Ref. No. K83133) and the Science and Technology Facilities Council
(STFC), UK for the financial and IDL support they received. The data is
provided courtesy of NASA/SDO and the AIA, EVE, and HMI science
teams. The STEREO/SECCHI data used here was produced by an
international consortium of the Naval Research Lab, Lockheed Martin
Solar and Astrophysics Lab, NASA Goddard Space Flight Center (USA),
Rutherford Appleton Laboratory , University of Birmingham (UK), Max-
Planck- Institut for Solar System Research (Germany), Center
Spatiale de Li\'{e}ge (Belgium), Institut d'Optique Th\'{e}orique et
Appliqu\'{e}e and Institut d'Astrophysique Spatiale (France).
\end{acknowledgements}


\begin{thebibliography}{}
\bibitem[Andries et al. (2009)]{ANDetal2009} {Andries}, J.,{van Doorsselaere}, T.,{Roberts}, B. {Verth}, G., {Verwichte}, E. \& {Erd{\'e}lyi}, R 2009, \ssr, 149, 3

\bibitem[Aschwanden (2004)]{ASC2004} Aschwanden, M.J. 2004, Physics of the Solar Corona (Berlin: Springer-Verlag)



\bibitem[Banerjee et al. (2007)]{BANETAL2007} {Banerjee}, D., {Erd{\'e}lyi}, R., {Oliver}, R. \& {O'Shea}, E 2007, \solphys, 246, 3

\bibitem[Bender \& Orszag (1978)]{BENORS1978} {Bender}, C.~M. \& {Orszag}, S.~A. 1978, Advanced Mathematical Methods for Scientists and Engineers, (New York: McGraw-Hill)

\bibitem[Bohlin et al. (1975)]{BOHetal1975} {Bohlin}, J.~D., {Vogel}, S.~N., {Purcell}, J.~D.,
    {Sheeley}, Jr., N.~R., {Tousey}, R. \& {Vanhoosier}, M.~E. 1975, \apj, 197, L133

\bibitem[Brueckner \& Bartoe (1983)]{BRUBAR1983} {Brueckner}, G.~E. and {Bartoe}, {J.-D.~F.}, \apj, 272, 329

\bibitem[Centeno et al. (2010)]{CENetal2010} {Centeno}, R., {Trujillo Bueno}, J. \& {Asensio Ramos}, A 2010, \apj, 708, 1579

\bibitem[Cirtain et al. (2007)]{CIRetal2007} Cirtain, J.W. et al. 2007, Science, 318, 1580

\bibitem[De Moortel (2009)]{DEM2009} {De Moortel}, I. 2009, \ssr, 149, 65

\bibitem[Del Zanna \& Mason (2003)]{DELMAS2003} {Del Zanna}, G. and {Mason}, H.~E. 2003, \aap, 406, 1089

\bibitem[Erd\'{e}lyi(1998)]{ERD1998} Erd\'{e}lyi, R.\ 1998, \solphys, 180, 213

\bibitem[Erd\'{e}lyi (2006a)]{ERD2006b} Erd\'{e}lyi, R. 2006a, Royal Society of London Philosophical Transactions Series A, 364, 351

\bibitem[Erd\'{e}lyi (2006b)]{ERD2006}  Erd\'{e}lyi, R. 2006b, Proceedings of SOHO 18/GONG 2006/HELAS I, Beyond the spherical Sun, ed. K. {Fletcher}, K. \& M. {Thompson}, ESA SP-624, p15

\bibitem[Erd\'{e}lyi et al.(2011)]{ERDetal2011} Erd\'{e}lyi, R., Al-Ghafri, K.~S., \& Morton, R.~J. 2011, \solphys, in press

\bibitem[Erd\'{e}lyi \& Fedun (2007)]{ERDFED2007} {Erd{\'e}lyi}, R. \& {Fedun}, V. 2007, Science, 318, 1572

\bibitem[Gabriel (1976)]{GAB1976} {Gabriel}, A.~H. 1976, Royal Society of London Philosophical Transactions Series A, 281, 339

\bibitem[Goossens et al.(2011)]{GOOSSetal2011} Goossens, M., Erd{\'e}lyi, R., \& Ruderman, M.~S. 2011, \ssr, in press

\bibitem[Harrison(1999)]{HARR1999} Harrison, R.~A. 1999, Magnetic Fields and Solar Processes, ESA SP-448, p531

\bibitem[Hollweg et al. (1982)]{HOLetal1982} {Hollweg}, J.~V. and {Jackson}, S. and {Galloway}, D. 1982, \solphys, 75, 35

\bibitem[Howard et al. (2008)]{HOWetal2008} Howard, R.A. et al. 2008, \ssr

\bibitem[Kamio et al. (2010)]{KAMetal2010} {Kamio}, S., {Curdt}, W., {Teriaca}, L., {Inhester}, B. \& {Solanki}, S.~K. 2010, \aap, 510, L1

\bibitem[Lemen et al. (2011)]{LEMetal2011} {Lemen}, J.~R., {Title}, A.~M., {Akin}, D.~J., {Boerner}, P.~F.,  {Chou}, C. {et al.} 2011, submitted

\bibitem[Liu et al. (2009)]{LIUetal2009} {Liu}, W., {Berger}, T.~E., {Title}, A.~M. \& {Tarbell}, T.~D. 2009, \apj, 707, L37

\bibitem[Liu et al. (2011)]{LIUetal2011} {Liu}, W., {Berger}, T.~E., {Title}, A.~M., {Tarbell}, T.~D. \&
    {Low}, B.~C. 2011, \apj, 728, 103

\bibitem[Morton \& Erd\'{e}lyi (2009a)]{MORERD2009} {Morton}, R. \& {Erd{\'e}lyi}, R., 2009a, \aap, 605, 493

\bibitem[Morton \& Erd\'{e}lyi (2009b)]{MORERD2009b} {Morton}, R. \& {Erd{\'e}lyi}, R., 2009b, \apj, 707, 750

\bibitem[Morton \& Erd\'{e}lyi (2010)]{MORERD2010} {Morton}, R.~J. \& {Erd{\'e}lyi}, R., \aap, 519, 43

\bibitem[Morton et al. (2010)]{MORHOOERD2010} {Morton}, R.~J., {Hood}, A.~W. \& {Erd{\'e}lyi}, R. 2010, \aap, 512, 23

\bibitem[Murawski \& Zaqarashvili (2010)]{MURZAQ2010} {Murawski}, K. \& {Zaqarashvili}, T.~V. 2010, \aap, 519, 8

\bibitem[Newton (1934)]{NEW1934} {Newton}, H.~W. 1934, \mnras, 94, 472

\bibitem[Pariat et al. (2009)]{PARetal2009} {Pariat}, E., {Antiochos}, S.~K. \& {DeVore}, C.~R. 2009, \apj, 691, 61

\bibitem[Patsourakos et al. (2008)]{PATetal2008} {Patsourakos}, S., {Pariat}, E., {Vourlidas}, A., {Antiochos}, S.~K. \&
    {Wuelser}, J.~P. 2008, \apj, 680, L73

\bibitem[Peter (2001)]{PET2001} {Peter}, H. 2001, \aap, 374, 1108

\bibitem[Pike \& Mason (1998)]{PIKMAS1998} {Pike}, C.~D. \& {Mason}, H.~E. 1998, \solphys, 182, 333

\bibitem[Pint{\'e}r et al. (2008)]{PINetal2008} {Pint{\'e}r}, B., {Jain}, R., {Tripathi}, D. \& {Isobe}, H. 2008, \apj, 680, 1560

\bibitem[Roberts et al. (1984)]{ROBetal1984} {Roberts}, B. and {Edwin}, P.~M. and {Benz}, A.~O. 1984, \apj, 279, 857

\bibitem[Ruderman (2010)]{RUD2011} Ruderman, M.S. 2010, \solphys, 267, 377

\bibitem[Ruderman \& Erd\'{e}lyi (2009)]{RUDERD2009} {Ruderman}, M.~S. \& {Erd{\'e}lyi}, R. 2009, \ssr, 149, 199

\bibitem[Ruderman et al. (2008)]{RUDetal2008} {Ruderman}, M.~S., {Verth}, G. \& {Erd{\'e}lyi}, R. 2008, \apj, 686, 694

\bibitem[Shibata et al. (1992)]{SHIetal1992} Shibata, K. et al. 1992, \pasj, 44, L173

\bibitem[Shibata \& Uchida (1985)]{SHIUCH1985} {Shibata}, K. \& {Uchida}, Y. 1985, \pasj, 37, 31

\bibitem[Shimojo et al. (1996)]{SHIMetal1996} {Shimojo}, M., {Hashimoto}, S., {Shibata}, K., {Hirayama}, T.,
    {Hudson}, H.~S. \& {Acton}, L.~W. 1996, \pasj, 48, 123

\bibitem[Soler et al. (2010)]{SOLetal2010} {Soler}, R., {Arregui}, I., {Oliver}, R. \& {Ballester}, J.~L. 2010, \apj, 722, 1778

\bibitem[Soler et al. (2009a)]{SOLetal2009} {Soler}, R., {Oliver}, R. \& {Ballester}, J.~L. 2009a, \apj, 699, 1553

\bibitem[Soler et al. (2009b)]{SOLetal2009b} {Soler}, R., {Oliver}, R. \& {Ballester}, J.~L. 2009b, \apj, 707, 662

\bibitem[Sterling et al. (2010)]{STEetal2010} {Sterling}, A.~C., {Harra}, L.~K. \& {Moore}, R.~L. 2010, \apj, 722, 1644

\bibitem[Tan et al. (2010)]{TANetal2010} {Tan}, B. and {Tian}, H. and {He}, J.~S. 2010, Chinese Astron. Astrophys., 34, 40

\bibitem[Taroyan \& Bradshaw (2008)]{TARBRA2008} {Taroyan}, Y. \& {Bradshaw}, S. 2008, \aap, 481, 247

\bibitem[Taroyan \& Erd\'{e}lyi (2009)]{TARERD2009} {Taroyan}, Y. \& {Erd{\'e}lyi}, R. 2009, \ssr, 149, 229

\bibitem[Terradas et al. (2011)]{TERetal2011} {Terradas}, J., {Arregui}, I., {Verth}, G. \& {Goossens}, M. 2011, \apj, 729, L22

\bibitem[Terradas et al. (2010)]{TERetal2010c} {Terradas}, J., {Goossens}, M. \& {Verth}, G. 2010, \aap, 524, 23

\bibitem[Uchida (1970)]{UCHIDA1970} Uchida, Y. 1970, \pasj, 22, 341

\bibitem[Van Doorsselaere et al. (2008)]{VANetal2008b} {Van Doorsselaere}, T., {Brady}, C.~S., {Verwichte}, E. \&
    {Nakariakov}, V.~M. 2008, \aap, 491, L9

\bibitem[Vasheghani Farahani et al. (2009)]{VAGetal2009} {Vasheghani Farahani}, S., {Van Doorsselaere}, T., {Verwichte}, E. \&
    {Nakariakov}, V.~M. 2009, \aap, 498, L29

\bibitem[Verth \& Erd\'{e}lyi (2008)]{VERERD2008} {Verth}, G. \& {Erd{\'e}lyi}, R. 2008, \aap, 486, 1015

\bibitem[Verth et al. (2008)]{VERERDJES2008} {Verth}, G., {Erd{\'e}lyi}, R. \& {Jess}, D.~B, 2008, \apj, 687, L45

\bibitem[Verth et al. (2011)]{VERetal2011} {Verth}, G. and {Goossens}, M. \& {He}, J. -S. 2011, \apj, 733, L15

\bibitem[Verth et al. (2010)]{VERTHetal2010} {Verth}, G., {Terradas}, J. \& {Goossens}, M. 2010, \apj, 718, L102

\bibitem[West et al. (2011)]{WESetal2011} {West}, M.~J., {Zhukov}, A.~N., {Dolla}, L. \& {Rodriguez}, L. 2011, \apj, 730, 122

\bibitem[Winebarger et al. (2001)]{WINetal2001} {Winebarger}, A.~R., {DeLuca}, E.~E. \& {Golub}, L. 2001, \apj, 553, L81

\bibitem[Yokoyama \& Shibata (1995)]{YOKSHI1995} {Yokoyama}, T. \& {Shibata}, K., 1995, \nat, 375, 42

\bibitem[Yokoyama \& Shibata (1996)]{YOKSHI1996} {Yokoyama}, T. \& {Shibata}, K., 1996, \pasj, 48, 353

\bibitem[Zaqarashvili \& Erd\'{e}lyi (2009)]{ZAQ2009} Zaqarashvili, T. V. \& Erd\'{e}lyi, R. 2009, \ssr, 149, 355

\end{thebibliography}
\end{document}